\input{aipcheck}

\documentclass[
    ,final            
  ]
  {aipproc}

\layoutstyle{6x9}

\begin{document}

\title{Neural networks with dynamical synapses: from mixed-mode oscillations and spindles to chaos}


\author{KyoungEun Lee}{
  address={Departamento de F{\'\i}sica da Universidade de Aveiro, I3N, 3810-193 Aveiro, Portugal}
}

\author{A. V. Goltsev}{
  address={Departamento de F{\'\i}sica da Universidade de Aveiro, I3N, 3810-193 Aveiro, Portugal}
   ,altaddress={A.F. Ioffe Physico-Technical Institute, 194021 St. Petersburg, Russia} 
}

\author{M. A. Lopes}{
  address={Departamento de F{\'\i}sica da Universidade de Aveiro, I3N, 3810-193 Aveiro, Portugal}
}

\author{J. F. F. Mendes}{
  address={Departamento de F{\'\i}sica da Universidade de Aveiro, I3N, 3810-193 Aveiro, Portugal}
}

\begin{abstract}
 Understanding of short-term synaptic depression (STSD) and other forms of synaptic plasticity
is a topical problem in neuroscience. Here we study the role of STSD in the formation of complex
patterns of brain rhythms. We use a cortical circuit model of neural
networks composed of irregular spiking excitatory and inhibitory neurons having type 1 and 2
excitability and stochastic dynamics. In the model, neurons form a sparsely connected network and
their spontaneous activity is driven by random spikes representing synaptic noise. Using simulations and analytical
calculations, we found that if the STSD is absent, the neural network shows either asynchronous behavior or 
regular network oscillations depending on the noise level. 
In networks with STSD, changing parameters of synaptic plasticity and the noise level, we observed
transitions to complex patters of collective activity: mixed-mode and spindle oscillations, bursts of
collective activity, and chaotic behaviour. Interestingly, these patterns are stable in a certain range of
the parameters and separated by critical boundaries. Thus, the parameters of synaptic plasticity can
play a role of control parameters or switchers between different network states. However, changes of
the parameters caused by a disease may lead to dramatic impairment of ongoing neural activity. We
analyze the chaotic neural activity by use of the 0-1 test for chaos (Gottwald, G. \& Melbourne, I., 2004) and show that it has a collective
nature.

\end{abstract}

\maketitle


\section{INTRODUCTION}

  Short-term synaptic depression (STSD) is an important form of short-term plasticity that 
provides a dynamic gain-control mechanism enhancing the sensitivity of cortical neurons to afferent firing patterns and 
expanding the range of possible coding strategies for cortical neurons \cite{Abbott97,Tsodyks97}.
Recent experimental studies and phenomenological model of STSD (so-called Tsodyks-Markram (TM) model, \cite{Tsodyks98}) have reported that synaptic transmitters are bombarded 
with the probability of release determined by the history of the synapses and by the arrival of the presynaptic spike (spike-timing dependent plasticity, STDP).
Hence, synaptic efficacy is changed and adapted according to the dynamics of presynaptic and postsynaptic neurons.
In turn, changes in synaptic efficacy influence activity of neurons.
Thus, interplay between STSD and neuronal activity is an underlying mechanism that influences collective dynamics of neural network, in particular, brain rhythms. 
At the present time, understanding of this influence is elusive. 

 The brain is always surrounded by noise and also noisy itself. Intuitively, noise is damaging. 
However, in neural networks, noise can play a positive role supporting oscillations and synchrony \cite{Ermentrout08,Faisal08}.
Ergo, stochastic processe is also one of the important ingredients of neural network.

 In the present work, we study the role of STSD in the formation of complex patterns of neuronal activities.
We use a cortical circuit model of stochastic neural networks having complex architecture and two types of neurons \cite{Goltsev10}. 
For describing STSD, the TM model is applied to excitatory-excitatory synapses. 
By simulations, we observed that STSD yields a rich reportoire of neuronal activities, such as mixed-mode oscillations (MMOs), spindles, and chaotic behaviour.
The parameters of synaptic plasticity can play a role of control parameters or switchers between different network states.
We analyze the chaotic neural activity by use of the 0-1 test for chaos \cite{Gottwald05,Gottwald09} and show that it has a collective nature.

\section{CORTICAL NETWORK MODEL}

\begin{figure}[tbhd]
\includegraphics[height=.3\textheight]{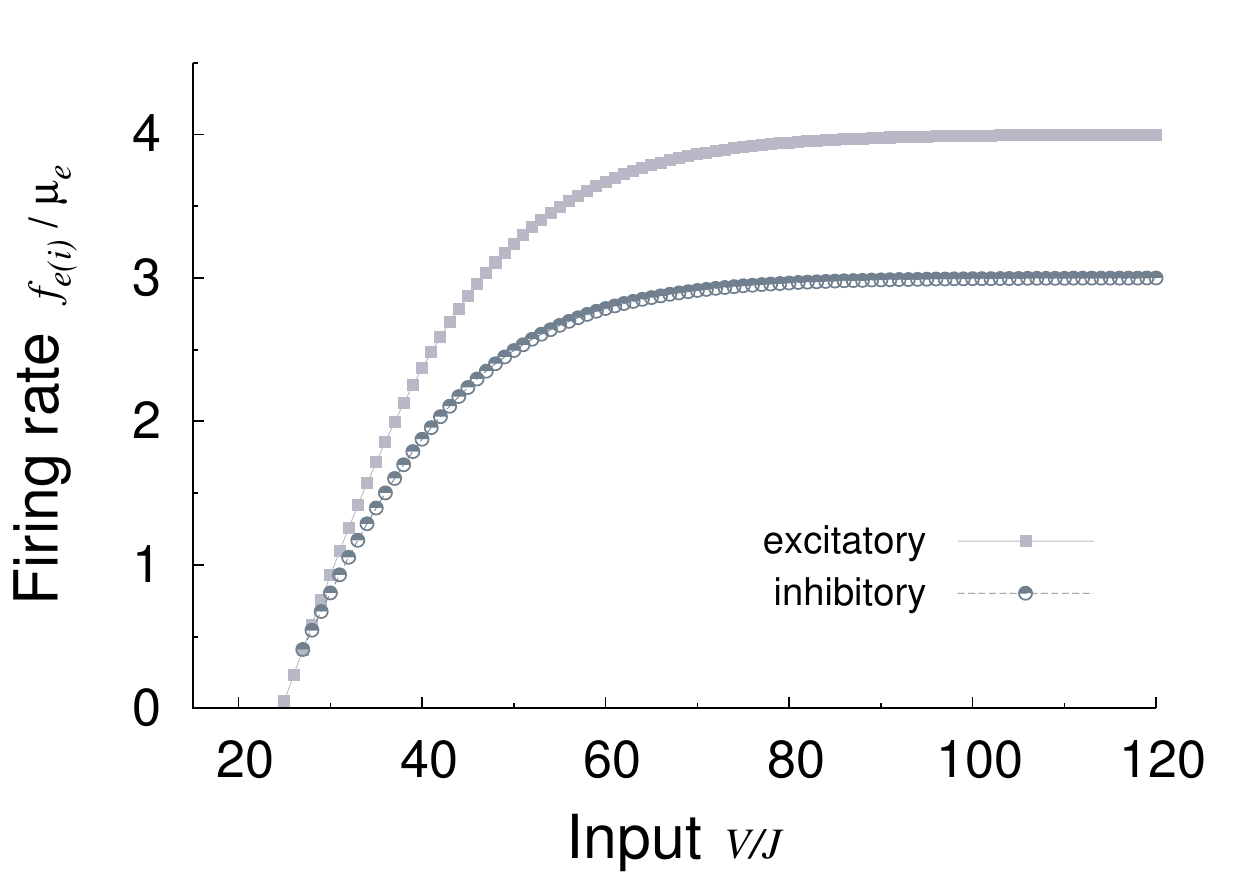}
\caption{Input-output function: firing rates (or spike frequencies) of excitatory and inhibitory neurons versus input $V$.
These firing rates follow sigmoid functions mimicking experimental reports.
Square points and solid lines are for excitatory neurons. Circle points and dashed lines are for inhibitory neurons.} \label{fig1}
\end{figure}

\emph{Stochastic neurons.  }
 Let us consider two types of neurons: excitatory and inhibitory neurons. The total number of neuron is $N$.
The fractions of excitatory and inhibitory neurons are $g_e$ and $g_i=1-g_e$, respectively. Neurons are linked by directed edges and form a network with an adjacency matrix $a_{nm}$,
where $m,n=1,2,...,N$. An element $a_{nm}$ is equal to $1$ if there is an edge directed from a neuron $m$ to a neuron $n$; otherwise, $a_{nm}=0$.
If a neuron $n$ is connected to a presynaptic neuron $m$, the synaptic efficacy in neuron $n$ is $J_{nm}(t)$.
We assume that all synapses of excitatory neurons are excitatory and all synapses of inhibitory neurons are inhibitory: this is the so-called Dale's principle \cite{Hoppensteadt97_book}.
Each neuron can be in either an active or inactive state.
We define $s_n(t)=1$ if neuron $n$ is active at moment $t$, and $s_n(t)=0$ if this neuron is inactive.
Active neurons fire random trains of spikes with a Poisson inter-spike intervals distribution \cite{Ermentrout08,Bedard06}.
We take no account of phase correlation between trains of random spikes generated by different neurons.

\emph{Network structure.  }
 We consider network structure is of a sparse random uncorrelated directed network, so-called directed Erd\"{o}s R\'{e}nyi graph. 
These networks are small worlds and can have an arbitrary degree distribution.
They are often considered as a good approximation to real networks \citep{Albert02,Dorogovtsev02,Newman03}. 
The advantage of these networks is that they can be studied analytically by use of mean-freld theory and easily modeled for simulation.

\emph{Noise.  }
 Each neuron is stimulated by Gaussian external noise with a mean rate in the range from $0$ to $20$ kHz. This is the only source of the noise in our model.
We assume that the noise is regarded as the signal from another brain area and also describes synaptic bombardment.
In simulations, we consider a Gaussian function of the probability distribution that a neuron obtains $\xi(t)$ random spikes via extrinsic stimuli each time interval $\tau$ (so-called integration time).

\emph{Input-output.  }
 Neurons demonstrate various types of spiking behaviour in response to a stimulus at firing threshold \cite{Izhikevich04,Izhikevich06}.
In particular, there are two types of threshold behaviour for periodcally firing neurons, which show continuous (type 1) or discontinuous (type 2) input-output curves \cite{Tateno04,Bogaard09}.
Excitatory neurons (pyramidal cells) represent type 1, whereas inhibitory neurons (interneurons) show type 2 behaviour.
In figure 1, we mimic the mean firing rates (in the present paper, we refer to as just \textquotedblleft firing rates \textquotedblright) described by sigmoid input-out functions of neurons reflecting the experimental reports.
At every integration time $\tau$, all neurons are updated in parallel, in which
postsynaptic neuron receives and integrates spikes from active presynaptic excitatory and inhibitory neurons and also Gaussian external input.

In a nutshell, the total number of input per neuron is:
    \begin{equation}
      V_m(t)=\sum_{n=1}^{N} k_n(t)a_{nm}J_{nm}+\xi(t),
    \end{equation}
where $k_n(t)$ is the number of spikes that arrive from presynaptic neuron $n$ and $<\xi>$ is the mean number of random spikes during the time interval $[t-\tau,t]$. 
Correspondingly, probabiliy distribution to get $k_{n}$ spikes from an active excitatory or inhibitory presynaptic neuron, $n$ obeys a Poisson distribution.
The inverse function of a mean firing rate indicates an average value of inter-spike interval times.	
Here we consider the case of $\tau f < 1$, that is, a postsynaptic neuron recieves only one spike ($k_{n}=1$) or none ($k_{n}=0$) from an active presynaptic neuron during the integration time $\tau$, 
thereby the parameter $\tau f$ has a meaning of the probability that a postsynaptic neuron receives a spike from an active presynaptic neuron during time $\tau$.

\emph{Activation-deactivation.  }
If the total input $V_m(t)$ at an inactive excitatory or inhibitory neuron $m$ at time $t$ is at least a certain threshold $\Omega_{a}$
 (i.e. $V_m(t) \geq \Omega_{a}$), then the neuron $m$ fires with a activation rate, $\mu_{a}$.
And the active excitatory (inhibitory) neuron $m$ is inactivated at a rate $\mu_a$ if $V_m(t) < \Omega_a$, provided that $a=e$ for excitatory and $a=i$ for inhibitory neurons, respectively.
The activation rate, $\mu_a$ is the reciprocal function of the mean first spike latency, $\Delta t_{sl}$.
Thus, activation of neurons also follows a stochastic process with a characteristic time ($t_r\equiv\Delta t_{sl}$).
For simplicity, we assume that activation and deactivation rate $\mu_a$ are the same and do not depend on the input.
The relation between excitatory and inhibitory activation rates is $\mu_i = \alpha \mu_e$ 

\paragraph{Short-term synaptic depression}
 The model of short-term synaptic depression is based on depletion of a pool of vesicles 
as a synaptic resource which is analogous to the probability of release in the phenomenological model described in (markram et al. 1998 \cite{Tsodyks98}).
It reflects the assumption that a releasable pool is determined not only by the history of the synapse 
but also by the arrival of the new presynaptic spike (STDP).
As a dynamical synapses, we only take synapses between excitatory neurons into account in the present work. Other synaptic efficacies are constant.
$J_{e}(t)=J_0y(t)$, $J_{0}$ is the absolute strength and $y(t)$ is a scaling factor which is the fraction of releasable synaptic resources.
The rate equation for evolution of $y$ is described by the following:

\begin{equation}
 \frac{dy}{dt} = \frac{1-y}{\tau_R} - P_d y\delta(t-t_s),
\end{equation}
where the control parameter $P_d$ and $\tau_R$ is the degree of depression (with $0 \leq P_d \leq 1$) and the recovery time from synaptic depression, respectively, 
and $t_s$ is the arrival timing of the presynaptic spikes,

The parameters can take any real values observed in experiments.
For the sake of convenience, we use dimensionless units for all parameters.
In simulations, the model parameters are set as followings: the network size $N=10^3-10^5$; 
the mean degree $10^3$; the fraction of inhibitory neurons $g_i=0.25$;
the all excitatory efficacies $J_e=1$ and all inhibitory efficacies $J_i=-3$;
the threshold $\Omega_e=25$, $\Omega_i=27$ for excitatory and inhibitory neurons, respectively;
the maximum mean firing rate of excitatory neurons $f_e^{max}=4$ and of inhibitory neurons$f_i^{max}=3$;
the activation rate of all excitatory neurons $\mu_e=1$; the deviation of the Gaussian random noise $\sigma=10$;
the ratio between first spike latency of excitatory and inhibitory neurons $\mu_i/\mu_e=\alpha=[0,1]$.
the fraction of available synaptic resources $P_d=[0,1]$.
These parameters agree with measurements in cortex.

\section{RESULTS}

In order to describe the dynamics of cortical neural networks, let us define $\rho_a(t)$ for excitatory, $a=e$, and inhibitory, $a=i$, neurons,
      \begin{equation}
	    \rho_a(t) \equiv \sum_{n} \rho_n^{(a)}(t)/(g_aN),
      \end{equation}
where $n$ is index of neuron and the sum is over neurons of type $a$, $g_a$ is their fraction. $\rho_e(t)$ and $\rho_i(t)$ are referred to as \textquotedblleft activities\textquotedblright of the excitatory and inhibitory populations.
In other words, $\rho_a(t)$ is the respective probabilities that a randomly chosen excitatory ($a=e$) or inhibitory ($a=i$) neuron is active at time $t$.
This activity is analogous to EEG measurements of brain waves.
To clarify the effects of depressing synapses, first we study a cortical network model in the absence of synaptic plasticity. 
This model has a complex phase diagram with self-organized top-down states, hybrid phase transitions, hysteresis phenomena, and a rich array of behaviours including decaying and stable oscillations, stochastic resonance, and neural avalanches.
It is also shown that global oscillations and stochastic resonance are intrinsic properties of this nonlinear dynamical system (also refer \cite{Goltsev10}).
By simulation, we found the phase diagram with three different regions on the $\alpha-<\xi>$ plane as shown in Fig. 2.
In neural networks without synaptic plasticity, neuronal activities, $\rho_a$ show different collective behaviours -- sustained I or suppressed states II, and global oscillations III depending on the parameter set, $\{ \alpha, <\xi> \}$.
The theory and simulations are in very good agreement at $N=100000$.
To investigate the effect of STSD on neural networks in the formation of complex patterns, we choose representative points on the phase II, III, and near the phase boundary between II and III (see the points A, B, C in Fig. 2).
The given points, A, B, C have the fixed average external stimulus, $<\xi> = 15$ and different control parameter $\alpha$ = $0.2$, $0.6$, $0.8$ for the oscillations, mixed states (II and III phases), and sustained active states, respectively.

\begin{figure}[tbhd]
\includegraphics[height=.3\textheight]{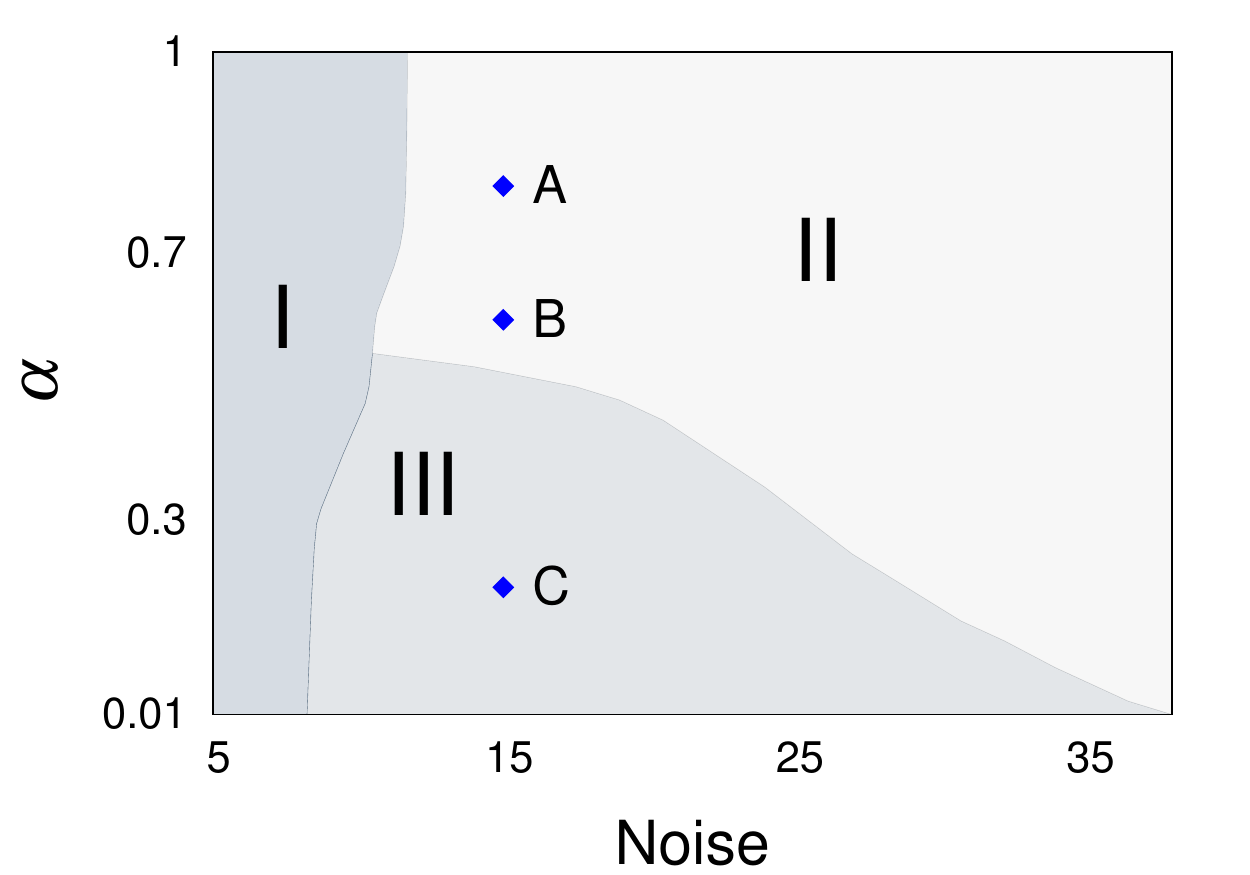}
\caption{Phase diagram, $\alpha$--$<\xi>$ plane: I. suppressed states, II. sustained active states, III. regular oscillations.
Points A, B, C show the parameters chosen for STSD; A. $\alpha=0.8$ and $<\xi>=15$, B. $\alpha=0.6$ and $<\xi>=15$, C. $\alpha=0.2$ and $<\xi>=15$.} \label{fig2}
\end{figure}

By STSD, neuronal activities turn out various brain rhythms which are experimentally observed in vitro and in vivo mammalian brain.
Through the given points A, B, C of the figure 2, in certain rages of synaptic parameters, we observed the new phases of neuronal activities which show chaotic behaviours.
Emergence of chaos is very intriguing phenomenon which is measured in EEG brain rhythms \citep{Babloyantz87, Freeman87}.

Regular oscillations of neuronal activity reveal amazing robustness against noise and depression of synapses in a broad range ($1-30$ Hz) of parameters.
However, we found that sufficiently strong synaptic bombardments (above $8$ Hz) stimulate synchronization between neurons and global neural oscillations when the synaptic depression is large enough.
With parameters $\alpha=0.2$ and $<\xi>=15$ (point C in the plane of $\alpha-<\xi>$), as depression parameter $P_d$ and/or recovery time $\tau_R$ are increased, new activity occurs to a certain extent in the intermediate region between regular oscillations and suppressed states.
The new activity has complex patterns that arise in neural networks, in which oscillations with different amplitudes are interspersed, 
that is, the neuronal activity switches between small amplitude oscillations and large amplitude oscillations.
This complex rhythm is reffered to as mixed-mode oscillations (MMOs). Figure 3 shows a sample snapshot of MMOs in time, where synaptic depression parameters are $P_d=0.01$, $\tau_R=51$.

\begin{figure}[tbhd]
\includegraphics[height=.3\textheight]{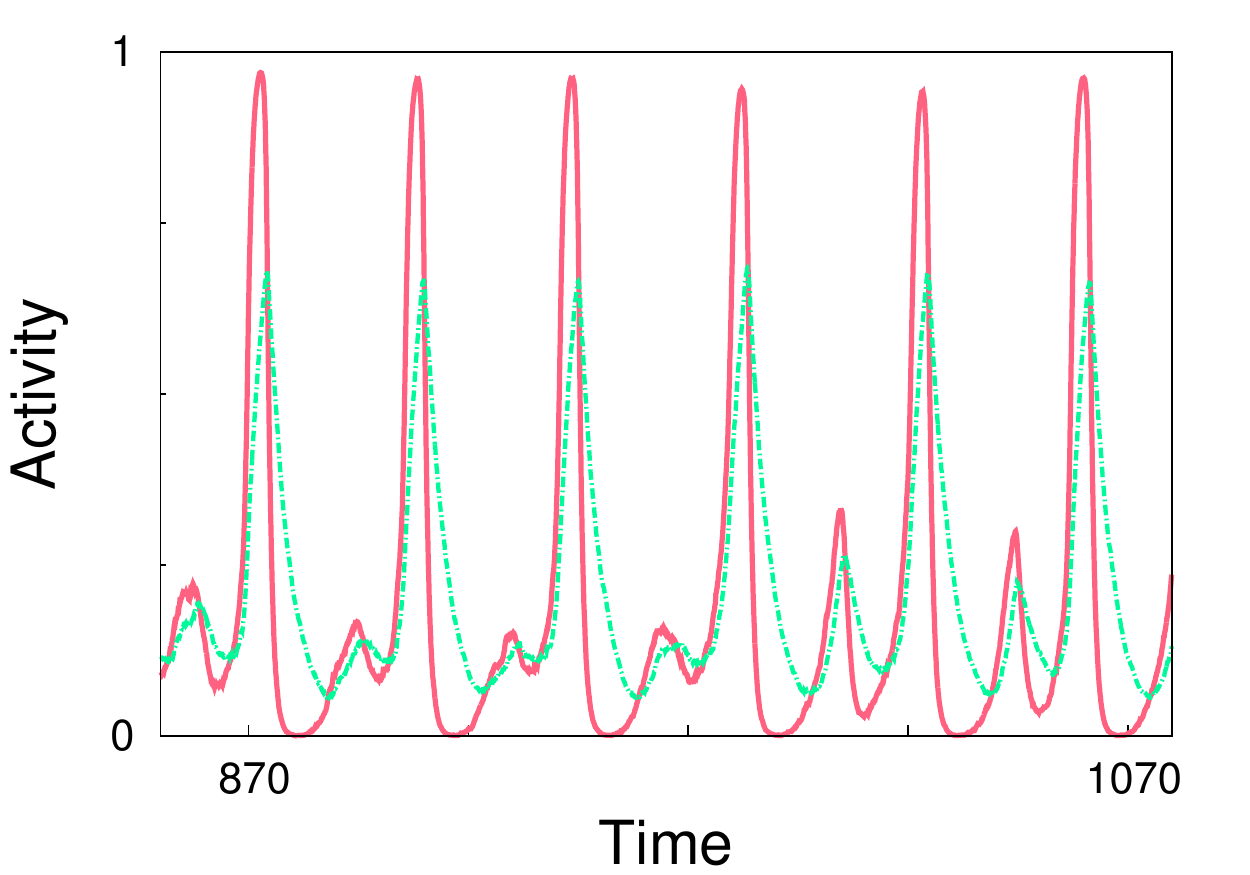}
\caption{Mixed-mode oscillations (or period-doubling oscillations): neuronal activity versus time. Parameters are $\alpha=0.2, <\xi>=15, P_d=0.01, \tau_R=51$.
Solid lines are for excitatory neurons, whereas dahsed-dotted lines are for inhibitory neurons.} \label{fig3}
\end{figure}


In the case of mixed state without plasticity (parameter $\alpha=0.6$, $<\xi>=15$, point B in Fig. 2), if neural network goes through STSD in certain range of parameters $\{P_d, \tau_R\}$, 
new phase of a so-called spindle-like oscillations appears.
Spindles are particularly generated during early stages of sleep. we observed these sleep spindles in our cortical model.
For example, with parameters $P_d=0.008$ and $\tau_R=9$, we found $4-5$ Hz spindle-like oscillations about $24$ spindles/min (see fig. 4-(b)).

\begin{figure}[h]
\includegraphics[height=.3\textheight]{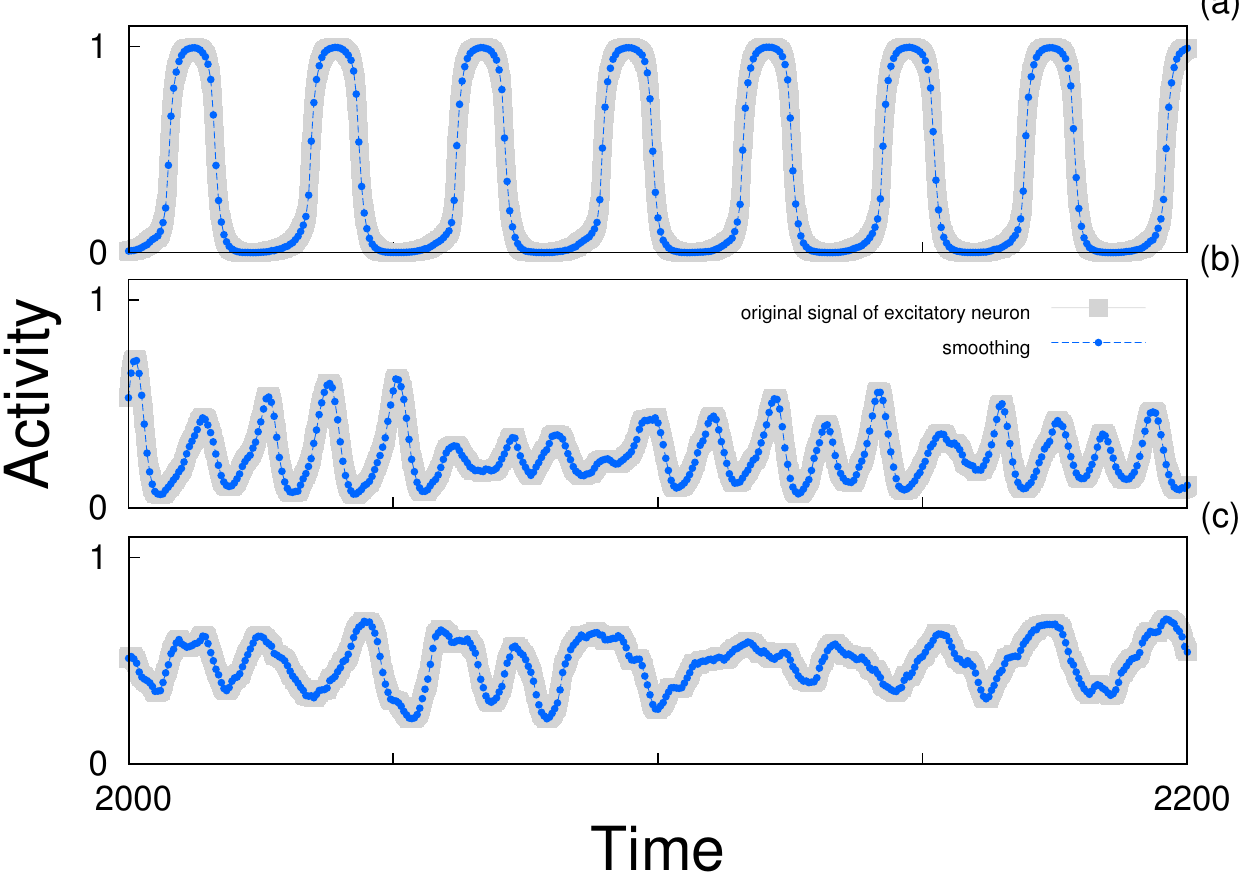}
\caption{the 0-1 test for chaos: neuronal activity vs. time. (a) $K_c=0.166197$ for regular oscillations ($\alpha=0.2$, $<\xi>=15$, $P_d=0.002$, $\tau_R=4$), 
(b) $K_c=0.764383$ for spindle oscillations ($\alpha = 0.6$, $<\xi>=15$, $P_d=0.008$, $\tau_R=9$), (c) $K_c=0.971950$ for chaotic oscillations ($\alpha=0.8$, $<\xi>=15$, $P_d=0.004$, $\tau_R=2.5$).
Square points and solid lines (gray color online) represent original signal of excitatory neurons 
while circle points and dashed lines (blue color online) show smoothing points (activities at every 5 integration times, $5\tau$).} \label{fig4}
\end{figure}


Lastly, when STSD is applied to initially sustained-active state (point A in Fig. 2, $\alpha=0.8$, $<\xi>=15$),
the neural network discloses another new phase in a certain range of parameters $\{P_d, \tau_R\}$.
The new phase exhibits chaotic activities as shown in Fig. 4-(c).
Typically MMOs and spindle oscillations have been also regarded as a chaotic behavour.
Here we make a narrower definition of chaotic behaviour. We primarily refer to chaotic activity as complex pattern that is lack of phase and frequency synchronization.


To check whether the observed activity in our model is chaotic or not, we use recent method, namely the 0-1 test for chaos introduced by Gottwald \& Melbourne \cite{Gottwald05,Gottwald09}. As $K_c \simeq 1$, the rhythm is wholly chaotic.
For a sample chaotic signal with the parameter set $\{\alpha=0.8, <\xi>=15, P_d=0.004, \tau_R=2.5\}$, we obtained $K_c = 0.971950$ (Fig. 4-(c)). thus, it is confirmed that our observed signals are chaotic.
Whereas, we acquired $K_c=0.764383$ from the data in the regime of spindles with the parameter set $\{\alpha=0.6, <\xi>=15, P_d=0.008, \tau_R=9\}$ (Fig. 4-(b))
and as would be expected, $K_c \simeq 0$ in the regular oscillation regime (Fig. 4-(a)).

\section{CONCLUSION}

By studying a cortical circuit model, we understand the role of STSD in the formation of complex patterns of brain rhythms.
Neuron is regarded as a generator of random spikes with a Poisson inter-spike intervals distribution and has type 1 (excitatory neuron )or type 2 (inhibitory neuron). 
Firing rate of excitatory (inhibitory) neuron is described by sigmoid function of input-out, 
And synaptic input from extrinsic afferents is modeled as an independent Gaussina process. 
Noise is a driving force and plays a crucial role in the dynamics of neural network. We use TM model \cite{Tsodyks98} for STSD of excitatory-excitatory dynamical synapses. 
By performing intensive simulations, we show that if the STSD is absent, the neural network exhibits either asynchronous behavior or regular network oscillations depending on the noise level 
(in agreement with analytical calculations and numerical analysis but with Heavy-side step function of firing rates).
Significantly, we also unravel the role of STSD in the present noise
whereby depressing excitatory synapses not only suppress neuronal activities but also on the contrary lead to a rich pattern of neural oscillations such as, mixed-mode oscillations, spindles, and chaotic activity.
We analyze the chaotic neural activity by use of the 0-1 test for chaos \cite{Gottwald05,Gottwald09} and show that it has a collective nature.
These patterns are stable in a certain range of the parameters and separated by critical boundaries.
Thus, the parameters of synaptic plasticity can play a role of control parameters or switchers between different network states. 
However, changes of the parameters caused by a disease may lead to dramatic dysfunction of ongoing neuronal activity.


\begin{theacknowledgments}
  This work was partially supported by the following projects PTDC:
SAU-NEU/ 103904/2008, FIS/108476/2008, MAT/114515/2009, and PEst-C/CTM/LA0025/2011.
K.~E.~Lee was supported by FCT under grant SFRH/BPD/ 71883/2010,
M.~A.~L. was supported by FCT under Grant No. SFRH/BD/ 68743/2010.

\end{theacknowledgments}



\bibliographystyle{aipproc}   

\bibliography{KELEE_proc}

\begin{thebibliography}{19}
\expandafter\ifx\csname natexlab\endcsname\relax\def\natexlab#1{#1}\fi
\providecommand{\enquote}[1]{``#1''}
\expandafter\ifx\csname url\endcsname\relax
  \def\url#1{\texttt{#1}}\fi
\expandafter\ifx\csname urlprefix\endcsname\relax\def\urlprefix{URL }\fi
\providecommand{\eprint}[2][]{\url{#2}}

\bibitem[Abbott et~al.(1997)]{Abbott97}
L.~F. Abbott, J.~A. Varela, K.~Sen, and S.~B. Nelson, \emph{Science}
  \textbf{275}, 221--224 (1997).

\bibitem[Tsodyks and Markram(1997)]{Tsodyks97}
M.~Tsodyks, and H.~Markram, \emph{Proceedings of the National Academy of
  Sciences} \textbf{94}, 719--723 (1997).

\bibitem[Tsodyks et~al.(1998)]{Tsodyks98}
M.~Tsodyks, K.~Pawelzik, and H.~Markram, \emph{Neural Computation} \textbf{10},
  821--835 (1998).

\bibitem[Ermentrout et~al.(2008)]{Ermentrout08}
G.~B. Ermentrout, R.~F. Gal\'an, and N.~N. Urban \textbf{31}, 428--434 (2008).

\bibitem[Faisal et~al.(2008)]{Faisal08}
A.~A. Faisal, L.~P.~J. Selen, and D.~M. Wolpert, \emph{Nat Rev Neurosci}
  \textbf{9}, 292--303 (2008).

\bibitem[Goltsev et~al.(2010)]{Goltsev10}
A.~V. Goltsev, F.~V. de~Abreu, S.~N. Dorogovtsev, and J.~F.~F. Mendes,
  \emph{Phys. Rev. E} \textbf{81}, 061921 (2010).

\bibitem[Gottwald and Melbourne(2005)]{Gottwald05}
G.~A. Gottwald, and I.~Melbourne, \emph{Physica D: Nonlinear Phenomena}
  \textbf{212}, 100 -- 110 (2005).

\bibitem[Gottwald and Melbourne(2009)]{Gottwald09}
G.~Gottwald, and I.~Melbourne, \emph{SIAM Journal on Applied Dynamical Systems}
  \textbf{8}, 129--145 (2009).

\bibitem[Hoppensteadt and Izhikevich(1997)]{Hoppensteadt97_book}
F.~Hoppensteadt, and E.~Izhikevich, \emph{Weakly Connected Neural Networks},
  Applied Mathematical Sciences V. 126, Springer, 1997.

\bibitem[B\'edard et~al.(2006)]{Bedard06}
C.~B\'edard, H.~Kr\"oger, and A.~Destexhe, \emph{Phys. Rev. Lett.} \textbf{97},
  118102 (2006).

\bibitem[Albert and Barab\'asi(2002)]{Albert02}
R.~Albert, and A.-L. Barab\'asi, \emph{Rev. Mod. Phys.} \textbf{74}, 47--97
  (2002).

\bibitem[Dorogovtsev and Mendes(2002)]{Dorogovtsev02}
S.~N. Dorogovtsev, and J.~F.~F. Mendes, \emph{Advances in Physics} \textbf{51},
  1079--1187 (2002).

\bibitem[Newman(2003)]{Newman03}
M.~Newman, \emph{SIAM Review} \textbf{45}, 167--256 (2003).

\bibitem[Izhikevich(2004)]{Izhikevich04}
E.~M. Izhikevich, \emph{IEEE transactions on neural networks / a publication of
  the IEEE Neural Networks Council} \textbf{15}, 1063--1070 (2004).

\bibitem[Izhikevich(2006)]{Izhikevich06}
E.~Izhikevich, \emph{Dynamical Systems in Neuroscience: The Geometry of
  Excitability and Bursting}, Computational Neuroscience, MIT Press, 2006.

\bibitem[Tateno et~al.(2004)]{Tateno04}
T.~Tateno, A.~Harsch, and H.~P.~C. Robinson, \emph{Journal of Neurophysiology}
  \textbf{92}, 2283--2294 (2004).

\bibitem[Bogaard et~al.(2009)]{Bogaard09}
A.~Bogaard, J.~Parent, M.~Zochowski, and V.~Booth, \emph{The Journal of
  Neuroscience} \textbf{29}, 1677--1687 (2009).

\bibitem[Babloyantz(1987)]{Babloyantz87}
A.~Babloyantz, \emph{Behavioral and Brain Sciences} \textbf{10}, 173--174
  (1987).

\bibitem[Skarda and Freeman(1987)]{Freeman87}
C.~A. Skarda, and W.~J. Freeman, \emph{Behavioral and Brain Sciences}
  \textbf{10}, 161--173 (1987).

\end{thebibliography}

\IfFileExists{\jobname.bbl}{}
 {\typeout{}
  \typeout{******************************************}
  \typeout{** Please run "bibtex \jobname" to optain}
  \typeout{** the bibliography and then re-run LaTeX}
  \typeout{** twice to fix the references!}
  \typeout{******************************************}
  \typeout{}
 }

\end{document}